\documentstyle[preprint,tighten,aps]{revtex}
\def\ut#1{\rlap{\lower1ex\hbox{$\sim$}}#1{}}
\begin{document}

\vspace*{2.5cm}
{\Large\bf Gauge invariance of complex general relativity}
%\vspace*{0.8cm}

{\bf Merced Montesinos\footnote{Department of Physics and Astronomy, 
University of Pittsburgh, Pittsburgh, PA 15260, USA. Permanent 
address: Departamento de F\'{\i}sica, Centro de Investigaci\'on y 
de Estudios Avanzados del I.P.N., Av. I.P.N. No. 2508, 07000 
Ciudad de M\'exico, M\'exico. e-mail: merced@fis.cinvestav.mx} 
and Jos\'e David Vergara\footnote{Instituto de Ciencias Nucleares, 
Universidad Nacional Aut\'onoma de M\'exico, 70-543, Ciudad de 
M\'exico, M\'exico. e-mail: vergara@nuclecu.unam.mx}} 

%\vspace*{0.5cm}
%\hspace*{1.5cm}
\begin{center}
\begin{minipage}[t]{9.5cm}\footnotesize 

\hrulefill

In this paper it is implemented how to make compatible the boundary 
conditions and the gauge fixing conditions for complex general relativity 
written in terms of Ashtekar variables using the approach of 
Ref. \cite{Henneaux1992}. Moreover, it is found that at first order in the 
gauge parameters, the Hamiltonian action is (on shell) fully 
gauge-invariant under the gauge symmetry generated by the first class 
constraints in the case when spacetime $\cal M$ has the topology 
${\cal M}= R \times \Sigma$ and $\Sigma$ has no boundary. Thus, the 
statement that the constraints linear in the momenta do not contribute to 
the boundary terms is right, but only in the case when $\Sigma$ has no 
boundary.

\hrulefill

\end{minipage}

\end{center}


It is a common belief that the (internal) gauge symmetry present in 
Yang-Mills theory is of a different kind with respect to the gauge symmetry 
involved in generally covariant theories as general relativity. This 
conclusion comes from the fact that when Yang-Mills theory is expressed in 
Hamiltonian form, the constraints are linear and homogeneous in their 
momenta. So, the Hamiltonian action of Yang-Mills theory is fully 
gauge-invariant under the gauge symmetry generated by the first class 
constraints. On the other hand, in covariant theories with first class 
constraints quadratic in their momenta, as general relativity expressed in 
terms of Arnowitt-Deser-Misner (ADM) variables, the Hamiltonian 
action is non gauge-invariant under the gauge symmetry generated by the 
first class constraints, rather, gauge invariance of the action is 
broken down [see for instance Ref. \cite{Teitelboim1982}]. The cause of 
this is precisely the quadratic (in the momenta) terms in the 
constraints. However, in Ref. \cite{Henneaux1992} was shown for generally 
covariant theories with a finite number of degrees of freedom that both 
type theories, those with constraints linear and homogeneous in the momenta 
and those with quadratic and higher order terms in the momenta, are very 
similar in some sense, the only difference (from the point of view of gauge 
symmetry) appears from the fact that in covariant theories is necessary to 
take into account more carefully the boundary terms. To handle the 
compatibility of the boundary conditions with the gauge fixing 
conditions, the approach developed in Ref. \cite{Henneaux1992} essentially 
takes advantage of the gauge symmetry generated by the gauge first class 
constraints.

In this paper, the programme of Ref. \cite{Henneaux1992} is implemented for 
self-dual gravity \cite{Ashtekar} where the first class constraints are 
quadratic in their momenta. It is very well known that complex general 
relativity can be rewritten in a Yang-Mills-like form if one uses Ashtekar 
variables to label the phase space of complex general relativity 
\cite{Ashtekar}, which opened an avenue for quantizing gravity from a 
nonperturbative viewpoint [see Refs.\cite{CarloI,CarloII} for a recent 
overview]. In spite of this success in the quantum theory, some facts still 
remain unclear even at classical level. For instance, a natural question is 
whether or not complex general relativity expressed in terms of Ashtekar 
variables shares the same property as `usual' Yang-Mills theory, namely, is 
the Hamiltonian action of complex general relativity in terms of Ashtekar 
variables fully gauge-invariant? Here, it is shown that the answer is in the 
affirmative. More precisely, the conclusion is that this Hamiltonian action 
is (on shell) gauge-invariant. It is important to mention that the 
techniques of Ref. \cite{Henneaux1992} had not been applied to a generally 
covariant theory with an infinite number of degrees of freedom, so this 
contribution fills out this gap in the literature. 

In what follows, it is assumed that the spacetime $\cal M$ has the 
topology ${\cal M} =R \times \Sigma$, and $\Sigma$ has no boundary, and 
Lorentzian signature is considered too. One starts with the 
self-dual (anti-self-dual) Hamiltonian form of complex general relativity 
expressed in terms of Ashtekar variables 
\begin{eqnarray}
S & = & \int^{x^0 =x^0_f}_{x^0=x^0_i} dx^0 \int_{\Sigma} d^3 x \left \{ 
{\dot A}^i_a {\widetilde \Pi}^a_i -
\left ( \ut{\lambda} {\widetilde{\widetilde H}} + 
{\lambda}^i {\widetilde G}_i + 
{\lambda}^a {\widetilde V}_a \right ) \right \}\, . \label{action}
\end{eqnarray}
The dependency of the Lagrange multipliers $\ut{\lambda}\,\,$, $\lambda^i$, 
$\lambda^a$, and the phase space variables $A^i_a$, ${\widetilde \Pi}^a_i$ 
on the Lagrangian variables is
\begin{eqnarray}
\ut{\lambda} & = & \frac{(N- N^b E^j_b M_j)}{E (1-M^2)} 
\left ( \frac{4\pi G}{c^3 q_{\epsilon}}\right )\, ,\nonumber\\
\lambda^i & = & - \frac{1}{q_{\epsilon}} \left ( 
\frac{1}{2} \epsilon^i\,_{jk} {^4\omega}_0\,^{jk}- 
\epsilon i {^4\omega}_0\,^{{\hat 0} i }\right ) \, , \nonumber\\
\lambda^a & = & N^a - 
\frac{(N- N^b E^j_b M_j)(E E^a_k M^k)}{E (1-M^2)} \, , \nonumber\\
A^i_a & = & \frac{1}{q_{\epsilon}} \left ( 
\frac{1}{2} \epsilon^i\,_{jk} {^4\omega}_a\,^{jk}-
\epsilon i {^4\omega}_a\,^{{\hat 0} i }\right ) \, , \nonumber\\
{\widetilde \Pi}^a_i & = &  \left (
\frac{c^3 i \epsilon q_{\epsilon}}{8\pi G}\right ) E (E^a_i + \epsilon i 
\epsilon_i\,^{jk} M_j E^a_k )\, .
\end{eqnarray}
$\epsilon=1$ ($\epsilon=-1$) corresponds with the self-dual 
(anti-self-dual) action. In the foliation it has been used the splitting of 
the (inverse) tetrad field $e^{\hat 0} = N dx^0 + M_i E^i_a dx^a$, 
$e^i = E^i_a N^a dx^0 + E^i_a dx^a$. Also, $E=\det{E^i_a}$, $E^a_i$ is the 
inverse of $E^i_a$; $M^2=M_i M^i$ with $M^i = M_j \delta^{ji}$. The 
constraints have the form 
\begin{eqnarray}
{\widetilde{\widetilde H}}
& = & \epsilon^{ijk} {\widetilde \Pi}^a_i
{\widetilde \Pi}^b_j F_{abk} (A) \, , \nonumber \\
{\widetilde G}_i & = & {\cal D}_a{\widetilde \Pi} ^a_i = \partial_a 
{\widetilde \Pi}^a_i - q_{\epsilon} \epsilon_{ij}\,^{k} A^j_a 
{\widetilde \Pi}^a_k \, ,\\
{\widetilde V}_a & = & {\widetilde \Pi}^b_i F_{ab}\,^i (A) \, ,  \label{GRC}
\end{eqnarray}
with $F_{ab}\,^i (A)= \partial_a A^i_b - \partial_b A^i_a - q_{\epsilon} 
\epsilon^{i}\,_{jk} A^j_a A^k_b$ the curvature of $A^i_a$. The symplectic 
structure of the phase space is given by
\begin{eqnarray}
\{ {\widetilde \Pi}^a_i (x)\, , \,{\widetilde \Pi}^b_j (y)\}
& = & 0\, ,\nonumber\\
\{ A^i_a (x) \, , \,{\widetilde \Pi}^b_j (y)\}
& = & \delta^b_a \delta^i_j \delta^3 (x,y)\, ,\nonumber\\
\{ A^i_a (x)\, , \, A^j_b (y)\}& = &  0 \, ,
\end{eqnarray}
where $a$ is a spatial index and runs over 1,2,3; the index $i$ runs over 
the Lie algebra of $SL(2,C)$. The self-dual (anti-self-dual) 
coupling constants are chosen as $q_{\epsilon}=-1$ as usual. Sometimes the 
`time gauge' is chosen too, $M_i=0$, which simplifies the definitions of 
the phase space variables. This can be done but it is not necessarily in 
principle. Up to here, it all has concerned with the first order 
formalism. It is possible to make contact with the second order 
formalism. Assuming there is no torsion and using 
$0=\nabla_{\alpha} e^I_{\beta}= 
\partial_{\alpha} e^I_{\beta} - {^4\omega}_{\alpha J}\,^I e^J_{\beta} + 
A_{\alpha\beta}\,^{\gamma} e^I_{\gamma}$ 
follows $\partial_{[a} E^i_{b]} - {^4\omega}_{[a \hat 0}\,^i E^j_{b]} M_j + 
\epsilon^i\,_{jk} {^4\Gamma}^j_{[a} E^k_{b]}$ with 
${^4\Gamma}^i_a = -\frac{1}{2} \epsilon^{i}\,_{jk} {^4\omega}_a\,^{jk}$ 
[see the definition of $A^i_a$]. When the `time gauge' is fixed, $M_i=0$, 
and then ${^4\Gamma}^i_a = \Gamma^i_a$ with $\Gamma^i_a$ given by 
$\partial_{[a} E^i_{b]} + \epsilon^i\,_{jk} {\Gamma}^j_{[a} E^k_{b]}=0$.

Due to the fact that the algebra of constraint closes with the 
diffeomorphism constraint ${\widetilde D}_a= \left (  
{\widetilde \Pi}^b_i F_{ab}\,^i - 
A^i_a {\widetilde G}_i \right )$ instead of the vector constraint 
${\widetilde V}_a$, and because one is interested in computing the effect 
of the gauge symmetry generated by the first class constraints on the 
action, it will be considered from now on the Hamiltonian action 
\begin{eqnarray}
S & = & \int^{x^0=x^0_f}_{x^0=x^0_i} dx^0 \int_{\Sigma} d^3 x \left \{ 
{\dot A}^i_a {\widetilde \Pi}^a_i - 
\left ( \ut{\lambda} {\widetilde{\widetilde H}} + 
{\lambda}^i {\widetilde G}_i + 
{\lambda}^a {\widetilde D}_a \right ) \right \}\, , \label{actionII}
\end{eqnarray}
instead of Eq. (\ref{action}).

The diffeomorphism constraint ${\widetilde D}_a$ as well as the Gauss 
constraint are linear in the momentum 
${\widetilde \Pi}^a_i$. Therefore, one can try to implement for 
gravity the ideas of Ref. \cite{Henneaux1992}. Due to the fact that the 
Hamiltonian constraint ${\widetilde{\widetilde H}}$ is quadratic in the 
momenta ${\widetilde \Pi}^a_i$, it could be expected that a 
(temporal) boundary term should be added to the action of Eq. 
(\ref{actionII})  to get back a fully covariant action for complex general 
relativity. Following Refs. \cite{Teitelboim1982,Henneaux1992}, the change 
on the action of Eq. (\ref{actionII}) generated by the Hamiltonian, 
diffeomorphism and Gauss constraints is to order $\varepsilon$
\begin{eqnarray}
S' & = & 
S [A' , {\widetilde \Pi}' , \ut{\lambda}' , {\lambda'}^i , {\lambda'}^a ] 
= S [A , {\widetilde \Pi}, \ut{\lambda} , {\lambda}^i , {\lambda}^a ] + 
\delta_{\varepsilon} B \mid^{x^0=x^0_f}_{x^0=x^0_i} \, , 
\end{eqnarray}
with
\begin{eqnarray}
{A'}^i_a & = & A^i_a + 2 \ut{\varepsilon}\, \, \epsilon^{ijk} 
{\widetilde \Pi}^b_j F_{abk} - {\cal D}_a \varepsilon^i 
+ {\cal L}_{\vec{\varepsilon}} A^i_a \, ,\nonumber\\
{\widetilde \Pi}^{'a}_i & = & {\widetilde \Pi}^a_i + 2 \left [ 
\partial_b \left ( \ut{\varepsilon}\, \,\epsilon_i\, ^{jk} 
{\widetilde \Pi}^b_j {\widetilde \Pi}^a_k \right ) + \epsilon_{im}\,^n A^m_b 
\left ( \ut{\varepsilon}\,\, \epsilon_n\,^{jk} {\widetilde \Pi}^b_j 
{\widetilde \Pi}^a_k \right ) \right ] \nonumber\\
& & + \epsilon_{ij}\,^k \varepsilon^j {\widetilde \Pi}^a_k + 
{\cal L}_{\vec{\varepsilon}} {\widetilde \Pi}^a_i \, , \nonumber\\
\ut{\lambda'} & = & \ut{\lambda} + \dot{\ut{\varepsilon}} + 
{\cal L}_{\vec{\varepsilon}}\, \ut{\lambda} - 
{\cal L}_{\vec{\lambda}}\, \ut{\varepsilon}  \,\,\,\, , 
\nonumber\\
{\lambda'}^i & = & \lambda^i + {\dot{\varepsilon}}^i + 4 
{\widetilde \Pi}^a_j {\widetilde \Pi}^{bj} ( \ut{\lambda}\,
\partial_b \ut{\varepsilon} - \ut{\varepsilon} \, \partial_b \ut{\lambda} ) 
A^i_a - [ \lambda , \varepsilon ]^i + 
{\cal L}_{\vec{\varepsilon}} \lambda^i - 
{\cal L}_{\vec{\lambda}} \varepsilon^i \, ,\nonumber\\
{\lambda'}^a & = & \lambda^a + {\dot{\varepsilon}}^a + 4 
{\widetilde \Pi}^a_j {\widetilde \Pi}^{bj} ( \ut{\lambda}\,
\partial_b \ut{\varepsilon} - \ut{\varepsilon} \, \partial_b \ut{\lambda} ) 
- [\lambda , \varepsilon ]^a \, ,\nonumber\\
\delta_{\varepsilon} B & = & \int d^3 x {\widetilde \Pi}^a_i (x)
\frac{\delta}{\delta {\widetilde \Pi}^a_i (x)} \int d^3 y \{
\ut{\varepsilon}\,\, {\widetilde{\widetilde H}}+
\varepsilon^i {\widetilde G}_i +
\varepsilon^a {\widetilde D}_a \}  \nonumber\\
& & - \int d^3 x \{
\ut{\varepsilon}\,\, {\widetilde{\widetilde H}}+ 
\varepsilon^i {\widetilde G}_i + 
\varepsilon^a {\widetilde D}_a \} \, . \label{gtrans}
\end{eqnarray}
Here, ${\cal L}_{\vec{\varepsilon}} A^i_a = \varepsilon^b \partial_b A^i_a 
+A^i_b \partial_a \varepsilon^b$, $ {\cal L}_{\vec{\varepsilon}} 
{\widetilde \Pi}^a_i = \varepsilon^b \partial_b {\widetilde \Pi}^a_i - 
{\widetilde \Pi}^b_i \partial_b \varepsilon^a + {\widetilde \Pi}^a_i 
\partial_b \varepsilon^b$, $[ \lambda , \varepsilon]^i = 
\epsilon^i\,_{jk} \lambda^j \varepsilon^k$, $[\lambda , \varepsilon ]^a = 
{\cal L}_{\vec{\lambda}} \varepsilon^a$. By inserting in 
$\delta_{\varepsilon} B$ the expressions  of the Hamiltonian, 
diffeomorphism, and Gauss constraints, it can be seen immediately that the 
contribution of both diffeomorphism and Gauss constraints is 
vanishing. Therefore, the boundary term $B$ is determined by the Hamiltonian 
constraint only, and is given by 
\begin{eqnarray}
\delta_{\varepsilon} B & = & \int d^3 x {\widetilde \Pi}^a_i (x) 
\frac{\delta}{\delta {\widetilde \Pi}^a_i (x)} \int d^3 y \,\, 
\ut{\varepsilon}\,\, {\widetilde{\widetilde H}} - \int d^3 x \,\,
\ut{\varepsilon}\,\, {\widetilde{\widetilde H}} = 
\int d^3 x \,\, \ut{\varepsilon} {\widetilde{\widetilde H}} \, .
\end{eqnarray}
Thus, 
\begin{eqnarray}
S' & = & 
S [A' , {\widetilde \Pi}' , \ut{\lambda}' , {\lambda'}^i , {\lambda'}^a ] 
= S [A , {\widetilde \Pi}, \ut{\lambda} , {\lambda}^i , {\lambda}^a ] + 
\int d^3 x \,\, \ut{\varepsilon} {\widetilde{\widetilde H}}
\mid^{x^0=x^0_f}_{x^0=x^0_i} \, . \label{gauge}
\end{eqnarray}
What is surprisingly is that $S'$ and $S$ differ by a (temporal) surface 
term involving the Hamiltonian constraint only. Now, the Dirac 
formalism \cite{Dirac} establishes first class constraints weakly 
vanish at any time $x^0$, which means the boundary term in 
Eq. (\ref{gauge}) vanishes `on shell'. Therefore, the action of 
Eq. (\ref{actionII}) is (on shell) fully gauge-invariant under the 
gauge symmetry generated by the first class 
constraints. Some comments are in order. First, general 
relativity in terms of Ashtekar variables shares the same property as 
standard Yang-Mills theory, in the sense that gauge invariance is not drop 
at the (temporal) boundary, {\it even though} complex general relativity 
has a constraint quadratic in their momenta. In Yang-Mills theory, on the 
other hand, the action is fully gauge-invariant and not only `on shell' due 
to the fact that the constraints are linear in the momenta. Whether or not 
this similarity between general relativity and Yang-Mills theory is a deep 
one deserves to be investigated. For the present purposes, it has been 
exhibited only that complex general relativity is truly `close' to a 
Yang-Mills theory in the sense explained. This property of the Hamiltonian 
action of complex general relativity in terms of Ashtekar variables is not 
present even in simple models with finite degrees of freedom, where $S'$ and 
$S$ differ by a (temporal) term which is {\it not} proportional to the 
quadratic constraints\footnote{For instance, in the case of the 
parameterized harmonic oscillator, the action 
$S = \int d\tau \{ \dot x p + \dot t p_t - \lambda (p_t + \frac{p^2}{2m} + 
\frac{1}{2} m \omega^2 x^2 ) \}$ is {\it not} fully gauge-invariant under 
the finite gauge transformations of the phase space variables and the 
Lagrange multiplier generated by the first class constraint 
$H= p_t + \frac{p^2}{2m} + \frac{1}{2} m \omega^2 x^2$, rather, under the 
gauge symmetry the action transforms as 
$S[ x' , t' , p' , {p'}_t , \lambda' ] = S[x, t, p, p_t , \lambda ] 
+ \left [ -\sin^2{\theta} x p + \frac{1}{\omega} \left ( 
\frac{p^2}{2m} - \frac{1}{2} m \omega^2 x^2 \right ) 
\frac{\sin{2\theta}}{2}\right ]^{\tau=\tau_2}_{\tau=\tau_1}$ 
\cite{Mon2000}. Still, it is always 
possible to build a fully covariant action by adding a suitable (temporal) 
boundary term. This new action is 
$S_{inv}=S -\frac{1}{2} x p \mid ^{\tau_f}_{\tau_i}$. Notice that the 
added boundary term is {\it not} proportional to the first class 
constraint $H$. In the ADM formalism of general relativity 
a similar result appears \cite{Teitelboim1982}.}.

Now, it is time to go to the issue concerning the compatibility of the 
boundary conditions and the gauge fixing conditions. For 
the action $S[A , {\widetilde \Pi}, \ut{\lambda} , {\lambda}^i , 
{\lambda}^a]$ the boundary conditions are
\begin{eqnarray}
A^i_a (x^0_i, x^a ) & = & {\cal A}^i_a (x^a) \, ,\nonumber\\
A^i_a (x^0_f, x^a ) & = & {\cal B}^i_a (x^a) \, , \label{XX}
\end{eqnarray}
where ${\cal A}^i_a (x^a)$, and ${\cal B}^i_a (x^a)$ are smooth 
configurations of the connection field at the end points. These boundary 
conditions have to 
be compatible with the gauge fixing conditions in order to have a 
well-defined dynamics. The gauge fixing conditions, in general, are 
expressions of the form
\begin{eqnarray}
\chi (A^i_a , {\widetilde \Pi}^a_i , x^0 , x^a ) & = & 0 \, , \nonumber\\
\chi_a (A^i_a , {\widetilde \Pi}^a_i , x^0 , x^a ) & = & 0 \, , \quad  
a =1,2,3 \, ,\nonumber\\
\chi_i  (A^i_a , {\widetilde \Pi}^a_i , x^0 , x^a ) & = & 0 \, , \quad 
i=1,2,3 \, .
\label{GAUGE}
\end{eqnarray}
Suppose, for the moment, they were {\it non} compatible with the boundary 
conditions (\ref{XX}). Under this assumption, dynamics of the gravitational 
field would be in trouble. How to make compatible both things? One 
possibility is to use the approach of Ref. \cite{Henneaux1992}. The idea 
developed in Ref. \cite{Henneaux1992} is to take advantage of the gauge 
symmetry 
generated by the first class constraints. Instead of considering the action 
$S[A , {\widetilde \Pi}, \ut{\lambda} , {\lambda}^i , {\lambda}^a]$, it has 
to be taken into account a {\it new} action $S_{inv}$, related to 
$S[A , {\widetilde \Pi}, \ut{\lambda} , {\lambda}^i , {\lambda}^a]$, 
through the gauge symmetry \cite{Henneaux1992}. This action is  
\begin{eqnarray}
S_{first \,\, order} 
[A , {\widetilde \Pi}, \ut{\lambda} , {\lambda}^i , {\lambda}^a , 
\ut{\varepsilon}] = 
S [A , {\widetilde \Pi}, \ut{\lambda} , {\lambda}^i , {\lambda}^a ] +  
\int d^3 x \,\, \ut{\varepsilon} {\widetilde{\widetilde H}}
\mid^{x^0=x^0_f}_{x^0=x^0_i} \, . \label{NEW}
\end{eqnarray}
Taking into account 
Eq. (\ref{gtrans}), the RHS of Eq. (\ref{NEW}) can be written as 
\begin{eqnarray}
S_{first\,\,order}
[A , {\widetilde \Pi}, \ut{\lambda} , {\lambda}^i , {\lambda}^a , 
\ut{\varepsilon} ] 
& = & S [A , {\widetilde \Pi}, \ut{\lambda} , {\lambda}^i , {\lambda}^a ] + 
\int d^3 x \,\, \ut{\varepsilon} {\widetilde{\widetilde H}}
\mid^{x^0=x^0_f}_{x^0=x^0_i} \, , \nonumber\\
& = & \int^{x^0=x^0_f}_{x^0=x^0_i} dx^0 \int_{\Sigma} d^3 x \left \{ 
{\dot {A'}}^i_a {\widetilde {\Pi'}}^a_i - 
\left ( \ut{\lambda'} {\widetilde{\widetilde {H'}}} + 
{\lambda'}^i {\widetilde {G'}}_i + 
{\lambda'}^a {\widetilde {D'}}_a \right ) \right \}\, , \nonumber\\
& = & S [A' , {\widetilde {\Pi'}}, \ut{\lambda'} , {\lambda'}^i , 
{\lambda'}^a ] \, . \label{QQQQ}
\end{eqnarray}
From the last line of Eq. (\ref{QQQQ}) the action $S_{first\,\,order} 
[A , {\widetilde \Pi}, \ut{\lambda} , {\lambda}^i , {\lambda}^a , 
\ut{\varepsilon} ]$ can be considered a functional of the gauge 
related phase space variables ${A'}^i_a$ and 
${\widetilde {\Pi'}}^a_i$ and of the 
Lagrange multipliers $\ut{\lambda}'$, ${\lambda'}^i$, and 
${\lambda'}^a$. Keeping this in mind it is pretty obvious that the 
boundary conditions for $S_{first\,\,order} 
[A , {\widetilde \Pi}, \ut{\lambda} , {\lambda}^i , {\lambda}^a , 
\ut{\varepsilon}]$ are 
\begin{eqnarray}
{A'}^i_a (x^0_i, x^a ) & = & {{\cal A}'}^i_a (x^a) \, ,\nonumber\\
{A'}^i_a (x^0_f, x^a ) & = & {{\cal B}'}^i_a (x^a) \, , \label{QQ}
\end{eqnarray}
where ${{\cal A}'}^i_a (x^a)$, and ${{\cal B}'}^i_a (x^a)$ are the initial 
(at $x^0_i$) and final (at $x^0_f$) configurations of the gauge related 
connection field ${A'}^i_a$. Instead of considering the variational 
principle defined by the action $S_{first\,\,order} 
[A , {\widetilde \Pi}, \ut{\lambda} , {\lambda}^i , {\lambda}^a , 
\ut{\varepsilon} ]$ in the form given by the last 
line in the RHS of Eq. (\ref{QQQQ}) together with the boundary conditions 
of Eq. (\ref{QQ}), it is possible to go back to the original set of 
variables. There, the phase space variables 
are $A^i_a$ and ${\widetilde \Pi}^a_i$, the action is given by 
Eq. (\ref{NEW}) and its boundary conditions, using Eq. (\ref{gtrans}), are 
\begin{eqnarray}
\left ( A^i_a + 2 \ut{\varepsilon}\, \, \epsilon^{ijk} 
{\widetilde \Pi}^b_j F_{abk} - {\cal D}_a \varepsilon^i 
+ {\cal L}_{\vec{\varepsilon}} A^i_a \right ) (x^0_i, x^a) 
& = & {{\cal A}'}^i_a (x^a) \, ,\nonumber\\
\left ( A^i_a + 2 \ut{\varepsilon}\, \, \epsilon^{ijk} 
{\widetilde \Pi}^b_j F_{abk} - {\cal D}_a \varepsilon^i 
+ {\cal L}_{\vec{\varepsilon}} A^i_a \right ) (x^0_f, x^a) 
& = & {{\cal B}'}^i_a (x^a) \, . \label{NEWXX}
\end{eqnarray}
This new variational principle, composed by the new action (\ref{NEW}) and 
the new boundary conditions (\ref{NEWXX}), has two peculiarities. First, 
the action $S_{first\,\,order}
[A , {\widetilde \Pi}, \ut{\lambda} , {\lambda}^i , {\lambda}^a , 
\ut{\varepsilon}]$ 
is `invariant' in the sense explained, namely, in terms of the gauge related 
variables the action $S_{first\,\,order}$ is the original action $S$ 
but written in terms of the gauge related variables [see the RHS of Eq. 
(\ref{QQQQ})]. Second, the new action (\ref{NEW}) also allows to make 
compatible the boundary conditions (\ref{NEWXX}) with the gauge fixing 
conditions (\ref{GAUGE}), through the gauge parameters involved there 
\cite{Henneaux1992}. To do this, it has to be plugged into the gauge 
conditions (\ref{GAUGE}) the expressions of the phase space 
variables $A^i_a$ and ${\widetilde\Pi}^a_i$ in terms of the gauge related 
variables ${A'}^i_a$, ${\widetilde {\Pi'}}^a_i$ as well as of the gauge 
parameters $\ut{\varepsilon}$ , $\varepsilon^i$, and $\varepsilon^a$. From 
these equations, one needs to solve for the gauge 
parameters $\ut{\varepsilon}$ , $\varepsilon^i$, and $\varepsilon^a$ in 
terms of the gauge related phase space variables ${A'}^i_a$, and 
${\widetilde{\Pi'}}^a_i$ and plug them into the RHS of Eq. (\ref{NEW}) 
and Eq. (\ref{NEWXX}). Of course, `on shell' there is no reason to plug 
these expressions for the gauge parameters into the RHS of Eq. (\ref{NEW}) 
because $S_{first\,\,order} = S$ `on shell'. Notice that in the boundary 
conditions it will appear ${A'}^i_a (x^0_i, x^a) [= {{\cal A}'}^i_a (x^a)]$, 
${A'}^i_a (x^0_f, x^a) [= {{\cal B}'}^i_a (x^a)]$, 
${\widetilde {\Pi'}}^a_i (x^0_i, x^a)$, and 
${\widetilde {\Pi'}}^a_i (x^0_f, x^a)$. They 
play the role of `parameters'. In addition, 
${{\cal A}'}^i_a (x^a) \neq {\cal A}^i_a (x^a)$ and 
${{\cal B}'}^i_a (x^a) \neq {\cal B}^i_a (x^a)$ in general. There is no 
reason why they should be the same expressions. As opposed to 
here, in Ref. \cite{Henneaux1992}, the equality among them was adopted.

In summary, the method of Ref. \cite{Henneaux1992} is totally systematic and 
can be applied to complex general relativity in terms of Ashtekar 
variables without any problems. The well-defined variational principle 
features i) an action defined by the RHS of (\ref{NEW}), ii) boundary 
conditions given by (\ref{NEWXX}) where the gauge parameters 
are in terms of the gauge related variables (`parameters') iii) these 
boundary conditions are compatible with the gauge fixing 
conditions (\ref{GAUGE}). So, the method of Ref. \cite{Henneaux1992} can 
also be applied to any generally 
covariant field theory with the obvious modifications required in theories 
with an infinite number of degrees of freedom, as in the case of general 
relativity studied here. Of course, the present analysis has been restricted 
to space times ${\cal M}$ having the topology ${\cal M} = R\times \Sigma$, 
where $\Sigma$ has no boundary. Before concluding, three more comments 
concerning Barbero variables, Husain-Kuchar's model, and spatial boundaries 
for $\Sigma$. First, what would have happened if one had used Barbero 
variables instead of Ashtekar variables? In terms of Barbero variables the 
Hamiltonian constraint is modified by a non polynomial contribution in the 
phase space variables \cite{Barbero1995}
\begin{eqnarray}
{\widetilde{\widetilde H}}
& = & \epsilon^{ijk} {\widetilde E}^a_i
{\widetilde E}^b_j F_{abk} (A) + 
\frac{2 (\beta^2 +1)}{\beta^2} {\widetilde E}^a_{[i}
{\widetilde E}^b_{j]} (A^i_a - \Gamma^i_a ) (A^j_b - \Gamma^j_b) 
\, . \nonumber\\
\end{eqnarray}
Obviously, in Barbero case, the difference between $S'$ and $S$ is 
{\it not} proportional to the Hamiltonian constraint, and the action $S$ is 
not `on shell' gauge-invariant in this case. Nevertheless, following Ref. 
\cite{Henneaux1992}, is possible to add a suitable boundary term to $S$ to 
build a fully gauge-invariant action. No specific proposal for this boundary 
term is presented in this paper, but it is possible to build it 
in principle. Second, it is important to notice that the Husain-Kuchar 
model, which is `close' to complex general relativity in terms of Ashtekar 
variables, is fully gauge-invariant under the gauge symmetry generated by 
the diffeomorphism and Gauss constraints in the case when ${\cal M}$ has 
the topology ${\cal M} = R\times \Sigma$ and $\Sigma$ has no boundary. 
This is so because in that model the Hamiltonian constraint is missing 
\cite{Husain}. Finally, the issue of the gauge invariance of complex general 
relativity when $\Sigma$ has boundaries deserves to be studied. Due to the 
fact that diffeomorphism constraint `moves' spatial boundaries, it is 
natural to expect to add a boundary term to get back a fully covariant 
action, namely 
\begin{eqnarray}
S_{inv} & = & S - \int_{\Sigma} d^3 x\,\, C ({\widetilde \Pi}^a_i , A^i_a ) 
\mid^{x^0 = x^0_f}_{x^0=x^0_i} - \int^{x^0=x^0_f}_{x^0=x^0_i} dx^0 
\int_{\partial \Sigma} I ({\widetilde \Pi}^a_i , A^i_a) \,\, dS \, .
\end{eqnarray}
The study of this issue is left for future work.

%%%%%%%%%%%%%%%%%%%%%%%%%%%%%%%%%%%%%%%%%%%%%%%%%%%%%%%%%%%%%%%%%%%%%%%%%%%%
\section*{acknowledgments}
%%%%%%%%%%%%%%%%%%%%%%%%%%%%%%%%%%%%%%%%%%%%%%%%%%%%%%%%%%%%%%%%%%%%%%%%%%%%
Warm thanks to Jerzy Pleba\'nski and Joseph Samuel for their 
valuable comments. MM thanks financial support provided by the 
{\it Sistema Nacional de Investigadores} (SNI) of the Secretar\'{\i}a de 
Educaci\'on P\'ublica (SEP) of Mexico. The summer stay of MM at the 
Department of Physics and Astronomy of the University of Pittsburgh, where 
this paper was finished, is supported by the {\it Mexican Academy of 
Sciences} and {\it The United States-Mexico Foundation for Science}. Also 
MM thanks all the members of the Department of Physics and Astronomy of 
the University of Pittsburgh for their warm hospitality. JDV is partially 
supported by grants DGAPA-UNAM IN100397 and CONACyT 32431-E. 
%%%%%%%%%%%%%%%%%%%%%%%%%%%%%%%%%%%%%%%%%%%%%%%%%%%%%%%%%%%%%%%%%%%%%%%%%%%% 

\end{document}